\documentclass[sigconf]{acmart}
\AtBeginDocument{%
  }

\copyrightyear{2026}
\acmYear{2026}
\setcopyright{cc}
\setcctype{by}
\acmConference[FSE Companion '26]{34th ACM Joint European Software Engineering Conference and Symposium on the Foundations of Software Engineering}{July 05--09, 2026}{Montreal, QC, Canada}
\acmBooktitle{34th ACM Joint European Software Engineering Conference and Symposium on the Foundations of Software Engineering (FSE Companion '26), July 05--09, 2026, Montreal, QC, Canada}
\acmDOI{10.1145/3803437.3805555}
\acmISBN{979-8-4007-2636-1/2026/07}

\begin{document}

\title{The Impact of Documentation on Test Engagement in Pull Requests in OSS}

\author{Teal Amore}
\orcid{0009-0003-6429-5821}
\affiliation{%
  \institution{Eastern Michigan University}
  \city{Ypsilanti}
  \state{Michigan}
  \country{USA}
}
\email{tamore@emich.edu}

\author{Nathan Berman}
\affiliation{%
  \institution{Eastern Michigan University}
  \city{Ypsilanti}
  \state{Michigan}
  \country{USA}
}
\email{nberman@emich.edu}

\author{Siyuan Jiang}
\affiliation{%
  \institution{Eastern Michigan University}
  \city{Ypsilanti}
  \state{Michigan}
  \country{USA}
}\email{sjiang1@emich.edu}

\renewcommand{\shortauthors}{Amore et al.}

\begin{abstract}
Automated testing is crucial for maintaining open-source software quality. However, motivating contributors to include tests for code changes remains a challenge. While existing interventions, such as code coverage metrics and reviewer feedback, are often reactive and applied only after a pull request is opened, this study investigates whether documentation on testing can serve as a proactive measure to encourage testing behavior.

In this work, we investigate the relationship between documentation on testing and contributor testing behavior. We introduce the \emph{Test Engagement Ratio} (TER) to help understand testing frequency. Using data from 160 OSS repositories, we analyze the relationship between documentation comprehensiveness and TER. Our results show a weak but statistically significant positive correlation ($\rho=0.36$, $p<0.001$), which strengthens to a moderate relationship ($\rho=0.44$) in repositories with higher pull request activity. Documentation categories such as How to Run Tests and How to Write Tests show the strongest correlation with testing engagement. Furthermore, TER is found to be moderately correlated ($\rho=0.52$, $p<0.001$) with Test Code Ratio, providing preliminary evidence of its validity. Our findings suggest that documentation on testing may be associated with increased testing engagement. Future work will explore causality, documentation quality at a granular level, and cross-repository exposure effects. 
\end{abstract}



\maketitle

\section{Introduction}

For open-source software (OSS) repositories, pull requests that contain tests are often highly valued and more likely to get merged \cite{li_opportunities_2022}. Despite this, motivating contributors to write tests remains challenging. Most interventions to promote testing, like code coverage metrics or reviewer comments, are reactive because they occur only after a contributor opens a pull request. These reactive approaches fail to consistently educate contributors about the project-specific testing standards. We investigate whether documentation on testing can serve as a proactive measure to encourage contributors to test more frequently. 

OSS repository maintainers often hold that automated testing is essential for maintaining quality, due to time limitations commonly associated with manual testing methods \cite{pham_creating_2013}. However developers often differ in their opinions about testing \cite{daka_survey_2014} \cite{masood_like_2022} \cite{santos_myths_2023}, which impacts their engagement with it. Research suggests that testing earlier in the development process can reduce the costs associated with bug fixing \cite{rani_shift-left_2023}, motivating exploration of proactive approaches. Documentation on testing may act as proactive guidance, yet no prior work has explored whether it correlates with testing behavior.

This study examines whether documentation on testing correlates with contributor testing behavior in open-source software. We build on Falcucci et al.'s dataset \cite{falcucci_what_2025}, which categorizes contribution documentation into various testing types like How to Run Tests, Unit Testing, and Integration Testing. The number of present categories provides a measure of documentation comprehensiveness. We propose a metric \emph{test engagement ratio} (TER) as a scalable proxy for testing behavior, measuring the proportion of production-code pull requests that also modify test files. We address the following research questions:

\begin{itemize}
    \item \textbf{RQ1.} Do contribution guidelines with more testing categories present correlate with a higher proportion of test-including pull requests?
    \item \textbf{RQ2.} What documentation categories or repository metadata correlate with test-including pull requests?
    \item \textbf{RQ3.} Does the proportion of test-including pull requests positively relate to standard test quality metrics?
\end{itemize}

\textbf{Summary of Findings}

Our analysis of 160 OSS repositories yields the following key results:

\begin{itemize}
    \item \textbf{RQ1.} We find a weak positive correlation between documentation comprehensiveness and TER ($\rho=0.36$, $p < 0.001$). This strengthens to a moderate correlation ($\rho=0.44$) in repositories with higher pull request activity.
    \item \textbf{RQ2.} Documentation on How to Run Tests ($\rho = 0.34$, $p < 0.001$) and How to Write Tests ($\rho = 0.36$, $p < 0.001$) shows the strongest correlation with TER. Repository popularity metrics (stars, forks, watchers) show no significant relationship. 
    \item \textbf{RQ3.} TER has a moderate positive correlation with Test Code Ratio ($\rho=0.52$, $p < 0.001$), providing preliminary evidence of its validity as a measure of testing activity.
\end{itemize}

\section{Methodology}

To investigate the relationship between documentation on testing and contributor testing behaviors, we analyzed the merged pull requests from the repositories in Falcucci et al. \cite{falcucci_what_2025}. To identify which pull requests interact with both production code and tests, we developed file classification heuristics. We then propose a testing frequency metric and conduct correlation analyses between this metric and testing categories that are present in documentation.

\subsection{File Classification}

We classified files as test or production code using path-based heuristics from Madeja et al. \cite{madeja_automating_2021} and extended it to include $spec$ files, which are common in JavaScript test files. 

\subsection{Test Engagement Ratio (TER)}

Calculating traditional testing metrics presents challenges when analyzing a large dataset of repositories. While code coverage is often measured by both developers and researchers, calculating such a metric is difficult due to the challenges of successfully building and running the tests for a repository \cite{aniche_why_2015}. Due to these challenges, we propose a new metric, \emph{Test Engagement Ratio} (TER), which enables us to measure whether a contributor has engaged in testing without having to successfully build or run the tests for each repository. We use the interaction with a test file as a proxy for engaging in testing behaviors. Unlike traditional metrics, TER does not capture whether tests are executed or meaningful, only whether contributors interact with test artifacts.

We define the \emph{Test Engagement Ratio} as:

\[
\text{Test Engagement Ratio = } \frac{\sum_{r\in R}I_{\text{both}}(r)}{\sum_{r\in R}I_{\text{prod}}(r)}
\]

where $R$ is the set of pull requests for a repository and the indicator functions are defined as:

\[
I_\text{prod}(r) =
    \begin{cases}
        1 & \text{if PR $r$ modifies a production file},\\
        0 & \text{otherwise}.
    \end{cases}
\]

\[
I_\text{both}(r) =
    \begin{cases}
        1 & \text{if PR $r$ modifies both production and test files},\\
        0 & \text{otherwise}.
    \end{cases} 
\]

In words, TER is the ratio of pull requests that modify both production and test files to those that modify production files, capturing the likelihood that contributors who change production code also modify tests. 

\subsection{Test Documentation Diversity Score}

To quantify the comprehensiveness of a repository's documentation on testing, we define the test documentation diversity score. This score is calculated as the number of testing categories present in the repository's contribution documentation. Each category that is present contributes one point to the score. The 10 possible categories are detailed in Table 2. The range of possible scores is from 0 to 10, where a higher score indicates more comprehensive documentation on testing. 

\subsection{Data Collection}

Using the file classification heuristics defined above, we developed a tool to gather all merged pull requests from January 1, 2025 through December 1, 2025. For each pull request, we recorded whether it modified a test file or a production code file. From this data, we calculated each repository's TER for use in our correlation analyses. 

We excluded repositories that had no merged pull requests that modified production code during the study period. These repositories are not relevant to our study, since they lack any active development that could be influenced by documentation on testing.

\subsection{Data Analysis}

\subsubsection{Correlation Analysis}

To investigate the relationship between documentation on testing comprehensiveness and contributor testing behavior, we performed a correlation analysis. We assess the association between each repository's test documentation diversity score and its computed TER using Spearman's rank correlation coefficient \cite{spearman_proof_1961}. Spearman's $\rho$ was selected due to the ordinal nature of the documentation score and the non-normal distribution of TER values. We report both the correlation coefficients and p-values to reflect the effect size and statistical significance. 

\subsubsection{Exploratory Analysis of Testing Categories and Metadata}

In addition to the primary analysis, we also conduct exploratory correlation analyses between the TER and the presence of each individual testing category in the documentation and repository metadata, such as stars, forks, and watchers. This analysis is intended to explore possible patterns and relationships that may inform future hypotheses rather than seeking to provide confirmatory evidence. As with the primary analysis, we use Spearman's rank correlation coefficient due to the expectation of skewed metadata and the ordinality of the presence of the testing categories. Given this analysis is exploratory, $p$-values are reported descriptively. 

\subsubsection{Metric Validation for Test-Including Pull Requests}

To establish the validity of TER as a measure of test engagement, we compare it with an established testing metric, the Test Code Ratio, which has been used in previous studies and is treated as a standard in test quality metrics \cite{fard_javascript_2017}\cite{pham_communicating_2015}\cite{sizilio_nery_empirical_2019}\cite{zaidman_studying_2011}. 

The Test Code Ratio is defined as:
\[
\text{Test Code Ratio = } \frac{NLOC_{\text{test}}}{NLOC_{\text{all}}}
\]

where NLOC is defined as non-empty lines of code.

We obtained the NLOC count using Lizard \cite{noauthor_terryinlizard_2026} for the test and production code to calculate the Test Code Ratio for each included repository. We examine the relationship between TER and Test Code Ratio using Spearman's $\rho$.

\section{Results}

This section presents the findings from our analysis of the 160 OSS repositories. We begin by describing the dataset characteristics and then report results for each of the research questions, assessing the correlation between documentation on testing and testing frequency (RQ1), exploratory relationships with documentation category (RQ2), and validating the Test Engagement Ratio metric (RQ3).

\subsection{Dataset Overview}

\begin{table*}[t]
    \centering
    \caption{Repository Statistics Overview}
    \label{tab:repo_stats}
    \begin{tabular}{lrrrrr}
        \toprule
        \textbf{Metric} & \textbf{Min} & \textbf{Max} & \textbf{Mean} & \textbf{Bottom quartile} & \textbf{Top quartile} \\
        \midrule
        Production Code PRs & 1 & 31301 & 746 & 7 & 356 \\
        Test-Including PRs & 0 & 15249 & 346 & 2 & 139 \\
        TER & 0 & 1 & 0.37 & 0.16 & 0.56 \\
        Production NLOC & 48 & 952931 & 63834 & 6405 & 72458 \\
        Test NLOC & 0 & 1237567 & 43651 & 1079 & 32706 \\
        Test Code Ratio & 0 & 0.88 & 0.31 & 0.03 & 0.58 \\
        \bottomrule
    \end{tabular}
\end{table*}

From the original 200 repositories in Falcucci et al.'s dataset, we excluded 40 repositories (14 were duplicates, 1 was deleted, and 25 had no pull requests modifying production code). This resulted in a dataset of 160 repositories for analysis. 

While the maximum value of the test documentation diversity score was 10, no repository was found to have all the testing categories present. Instead the maximum found in this study was a score of 8. The test documentation diversity score has a mean value of 3.9.

For the TER, the range of possible values is between 0 and 1. We found that 5 of the repositories reported a score of 1. Those repositories had 1-2 pull requests merged during the study period. We found the mean to be 0.37.

To validate our assumptions that the test documentation diversity score and the TER are nonparametric distributions, we decided to perform the Shapiro-Wilk test \cite{shapiro_analysis_1965} for each variable. The results of these tests confirmed all metrics are non-normal distributions ($p<0.001$), justifying our use of Spearman's $\rho$ for correlation analysis.

\subsection{RQ1: Correlation between documentation on testing and testing frequency}

RQ1 examines whether having more comprehensive documentation about testing correlates with a higher TER.

Utilizing Spearman's $\rho$, we observed a correlation between the TER and test documentation diversity score of $\rho = 0.36$ and $p < 0.001$. This suggests there is a weak but statistically significant monotonic relationship between these two variables. 

To assess robustness, we conducted a sensitivity analysis excluding repositories with fewer than five merged production-modifying pull requests (28 removed) to reduce instability from small denominators. In this subset of repositories, we observed a correlation of $\rho = 0.44$ and $p < 0.001$, suggesting a stronger relationship. The relationship appears more pronounced in repositories with sufficient pull request activity. 

Repositories in the top quartile of documentation diversity (scores 6-8) have an average TER of 0.51 while those in the bottom quartile (scores 0-2) have an average of 0.27. This suggests that repositories with more comprehensive documentation on testing see nearly twice the rate of test-including pull requests compared to those with minimal documentation. 

\subsection{RQ2: Exploratory correlations}

\begin{table}[h]
    \centering
    \caption{Correlation between Test Engagement Ratio and documentation on testing categories}
    \begin{tabular}{lccc}
        \hline
        Variables & Correlation ($\rho$) & 95\% CI  & n \\
        \hline
        Has documentation & 0.01 & [-0.15, 0.16] & 148 \\
        Has test documentation & 0.16 & [0.01, 0.31] & 130 \\
        How to run tests & 0.34 & [0.19, 0.48] & 111 \\
        How to write tests & 0.36 & [0.21, 0.49] & 50 \\
        Unit tests & 0.29 & [0.13, 0.43] & 93 \\
        Integration tests & 0.26 & [0.11, 0.41] & 27 \\
        End-to-end tests & 0.11 & [-0.05, 0.27] & 23 \\
        Mocks & 0.06 & [-0.10, 0.22] & 13 \\
        Code coverage & 0.19 & [0.03, 0.35] & 33 \\
        Best practices/testing tips & -0.02 & [-0.18, 0.14] & 10 \\
        \hline
    \end{tabular}
\end{table}

RQ2 examines which categories of documentation on testing and repository metadata factors correlate with TER. This exploratory analysis aims to identify patterns that could inform future hypotheses about which documentation elements are most associated with test engagement. 

\subsubsection{Documentation Category Correlations}

Table 2 presents correlations between TER and categories of documentation on testing. Of the 10 categories examined, documentation on How to Run Tests ($\rho = 0.34$, $p < 0.001$) and How to Write Tests ($\rho = 0.36$, $p < 0.001$) exhibited the strongest relationship to test engagement. We also found that Unit Testing ($\rho = 0.29$, $p < 0.001$), Integration Testing ($\rho = 0.26$, $p = 0.001$), and Code Coverage ($\rho = 0.19$, $p = 0.02$) documentation demonstrated significant positive correlations. 

Several categories showed minimal and non-significant relationships with test engagement, including End-to-end Tests ($\rho = 0.11$, $p = 0.17$), Mocks ($\rho = 0.06$, $p = 0.48$), and Best Practices/Testing Tips ($\rho = -0.02$, $p = 0.78$). However these categories were present in less than 20\% of the repositories. This limits our ability to draw conclusions about their effectiveness in influencing testing engagement.

\subsubsection{Repository Metadata Correlations}

We examined whether repository popularity influenced testing engagement by analyzing the correlation between TER and common GitHub popularity metrics (stars, forks, watchers). We found that these popularity metrics showed no correlation with TER ($|\rho| < 0.04$, $p > 0.64$). This suggests that contributor's testing engagement isn't related to the popularity of a repository.  

\subsection{RQ3: Test Engagement Ratio Metric Validation}

RQ3 examines the relationship between Test Engagement Ratio and Test Code Ratio to establish if our metric behaves similarly to an accepted, standard metric. We examine the relationship between the two metrics using Spearman's $\rho$ across 114 repositories. This subset reflects repositories where the Test Code Ratio could be reliably computed. The remaining 46 repositories were excluded due to challenges in automatically detecting test and production code file paths required by Lizard \cite{noauthor_terryinlizard_2026} for NLOC calculation. The observed positive correlation had values of $\rho = 0.52$ and $p < 0.001$. These findings suggest a moderate relationship. 

\section{Related Work}

We situate our work within three areas of prior research: OSS documentation, reactive approaches to encourage testing, and proactive approaches to encourage testing.

\subsection{OSS Documentation}

Prior work has examined various forms of OSS documentation. Aghajani et al. \cite{aghajani_software_2020} found that a majority of software developers found it helpful to have documentation that covers software testing and quality assurance. Beyond developer perceptions, research has also focused on the structure of documentation itself. Prana et al. \cite{prana_categorizing_2019} categorized the contents of README files into seven categories based on the information present focusing on high level themes. More specific to testing, Falcucci et al. \cite{falcucci_what_2025} identified 10 distinct categories found in documentation on testing.

\subsection{Reactive Approaches to Encourage Testing}

Due to repository maintainers' desires for well-tested repositories, there is a wide body of literature on how maintainers can motivate contributors to contribute tests alongside code changes. Continuous Integration checks used in conjunction with the Modern Code Review process have been found to improve the quality of OSS repositories \cite{vasilescu_quality_2015} \cite{zampetti_study_2019}. However these tools can cause problems due to the complexity of configuration and maintainability requiring manual human-driven overrides. Recently AI code review tools have been introduced to mitigate some of these challenges and can decrease the time to leave comments and often are as thorough as human reviewers \cite{vijayvergiya_ai-assisted_2024} \cite{naulty_bugdar_2025}.

\subsection{Proactive Approaches to Encourage Testing}

While there is limited direct research on proactive approaches to encourage testing, there exists adjacent literature that can provide a foundation. In a similar vein as documentation on testing, the presence of repository badges in documentation has been found to correlate with improved testing metrics \cite{trockman_adding_2018}. Establishing a strong testing culture can further encourage contributors to test \cite{pham_creating_2013}, though high contributor turnover makes communicating this culture to newcomers challenging.

Blanco et al. found that gamifying testing increased testing activities among students \cite{blanco_can_2023}. While not directly measuring testing activity, adding pre-commit hooks has been found to improve build stability in projects \cite{baum_leveraging_2015}, suggesting a shift toward earlier testing in the development workflow. Finally studies on OSS motivation indicate that contributors are often driven by recognition \cite{hars_working_2002}, implying that explicitly recognizing testing contributions may encourage increased testing activity. 

\section{Future Plans}

We plan to further explore documentation on testing and its impacts on testing, including:

\emph{Causality evaluation.} While this study establishes correlation, causality remains unproven. To explore causality, we propose a natural experiment using a difference-in-differences design applied to repositories that introduce documentation on testing. We will compare the TER before and after documentation changes in treatment repositories against matched control repositories (by size, language, and project type) without changes. By controlling for confounders such as project age and contributor experience, we aim to isolate the causal influence of documentation on testing in real-world OSS repositories. A key limitation of this study is that the observed correlation may reflect an underlying testing culture, where mature projects naturally develop both documentation and testing practices. The proposed difference-in-differences design would help address this confound.

\emph{Fine-grained analysis of documentation diversity.} Our current work measures the presence of testing categories in documentation, but not the quality or comprehensiveness within each category. Future studies will evaluate documentation on testing at a more granular level to understand what topics and depth are most valued by contributors. 

\emph{Exposure effect of quality documentation.} This study focuses solely on the impact of documentation on testing within a single repository. However, we don't know if exposure to documentation on testing can lead to improved testing engagement in other repositories. We plan to explore further if there exists a cross-repository exposure effect by tracking contributors across multiple projects. 

\emph{Contributor-level examination of Test Engagement Ratio.} In this study, we explored using TER to measure and compare repositories. Due to the flexibility of this metric, we could also use it to compare individual contributors. We plan to explore what types of contributors exhibit higher TER and what factors influence changes in individual testing behaviors over time.

\section{Conclusion}

We investigate how documentation on testing influences contributor testing behavior in open-source software. We propose Test Engagement Ratio (TER) to help understand testing frequency and find a positive correlation between documentation comprehensiveness and testing engagement across 160 repositories. While causality has not yet been established, this work provides the first empirical evidence connecting documentation on testing to contributor testing behaviors. 

\section{Data Availability}

In an effort to promote transparency and reproducibility, our artifacts are publicly available \cite{amore_impact_2026}.

\bibliographystyle{ACM-Reference-Format}
\bibliography{citations}

\end{document}